\documentclass[11pt,a4paper]{article}

\usepackage{graphicx}
\usepackage{psfrag}
\usepackage{amsmath}
\usepackage{SGpreprint}
\usepackage{algpseudocode}
\usepackage{algorithm}

\renewcommand{\thesection}{\Roman{section}}

\nologohead{  
         J.B. Glattfelder and S. Battiston: \\
         Backbone of complex networks of corporations: The flow of control \\ 
        Physical Review E {\bf 80} (2009)
      } 

\begin{document}

\begin{center}
  \textbf{\Large Backbone of complex networks of corporations: \\
   The flow of control}\\[5mm]
  \textbf{ J.B. Glattfelder and S. Battiston}\\
  Chair of Systems Design, ETH Zurich, Kreuzplatz 5, 8032 Zurich, Switzerland \\
\end{center} 

\begin{abstract}
  We present a methodology to extract the backbone of complex networks
  based on the weight and direction of links, as well as on
  nontopological properties of nodes. We show how the methodology can be
  applied in general to networks in which mass or energy is flowing along
  the links. In particular, the procedure enables us to address important
  questions in economics, namely, how control and wealth are structured and
  concentrated across national markets. We report on the first
  cross-country investigation of ownership networks, focusing on the
  stock markets of 48 countries around the world. On the one hand, our
  analysis confirms results expected on the basis of the literature on
  corporate control, namely, that in Anglo-Saxon countries control tends
  to be dispersed among numerous shareholders. On the other hand, it also
  reveals that in the same countries, control is found to be highly
  concentrated at the global level, namely, lying in the hands of very few
  important shareholders. Interestingly, the exact opposite is observed
  for European countries. These results have previously not been
  reported as they are not observable without the kind of network
  analysis developed here.

\end{abstract}

\section{Introduction}
\label{section:introduction}

The empirical analysis of real-world complex networks has revealed
unsuspected regularities such as scaling laws which are robust across
many domains, ranging from biology or computer systems to society and
economics
\cite{albert1999idw,pastorsatorras2001dac,newman2002rgm,garlaschelli2003usr}.
This has suggested that universal or at least generic mechanisms are at
work in the formation of many such networks. Tools and concepts from
statistical physics have been crucial for the achievement of these
findings \cite{DM03,caldarelli2007sfn}.

In the last years, in order to offer useful insights into more
detailed research questions, several studies have started taking into
account the specific meaning of the nodes and links in the various
domains the real-world networks pertain to \cite{barrat04a,GL04b}.
Three levels of analysis are possible. The lowest level corresponds to
a purely topological approach where the network is described by a
binary adjacency matrix. By taking weights \cite{barrat04a}, or
weights and direction \cite{Onnela05}, of the links into account, the
second level is defined. Only recent studies have started focusing on
the third level of detail, in which the nodes themselves are assigned
a degree of freedom, sometimes also called fitness. This is a
nontopological state variable which shapes the topology of the
network \cite{GL04b,GBCSC05,demasi2006fmi}.

The physics literature on complex economic networks has previously
focused on boards of directors \cite{newman2001rga, battiston2004spc},
market investments \cite{GBCSC05,BRZ05}, stock price correlations
\cite{bonanno2003tcb,onnela2003dmc}, and international trade
\cite{GL04,serrano2003twt,fagiolo2009wtw}. Here we instead present a
comprehensive cross-country analysis of 48 stock markets
world wide. Our first contribution is an algorithm able to identify
and extract the backbone in the networks of ownership relations among
firms: the core subnetwork where most of the value and control of the
market resides. Notably, we also provide a generalization of the method
applicable to networks in which weighted directed links and
nontopological properties of nodes play a role. In particular, the
method is relevant for networks in which there is a flow of mass (or
energy) along the links and one is interested in identifying the subset
of nodes where a given fraction of the mass of the system is flowing. The
growing interest in methods for extracting the backbone of complex
networks is witnessed by recent work in similar direction
\cite{serrano2009emb}.

Furthermore, given the economic context of the analyzed networks, we
contribute a model to estimate corporate control based on the
knowledge of the ownership ties. In order to identify the key players
according to their degree of control, we take the value of the market
capitalization of the listed companies (a good proxy for their size) to
be the nontopological state variable of nodes in the network. Our main
empirical results are in contrast with previously held views in the
economics literature \cite{davis2008nfc}, where a local distribution of
control was not suspected to systematically result in global
concentration of control and vice versa.

\section{Dataset}
\label{data}

We are able to employ a unique data set consisting of financial
information of listed companies in national stock markets. The ownership
network is given by the web of shareholding relations from and to such
companies. The analysis is constrained to 48 countries given in
\ref{app:countries}. The data are compiled from Bureau van Dijk's
ORBIS database\footnote{\texttt{http://www.bvdep.com/orbis.html}.}.  In
total, we analyze 24 877 corporations (or stocks) and 106 141 shareholding
entities who cannot be owned themselves (individuals, families,
cooperative societies, registered associations, foundations, public
authorities, etc.). Note that because the corporations can also appear as
shareholders, the network does not display a bipartite structure. The
stocks are connected through 545 896 ownership ties to their
shareholders. The database represents a snapshot of the ownership
relations at the beginning of 2007. The values for the market
capitalization, which is defined as the number of outstanding shares
times the firm's market price, are also from early 2007. These values
will serve as the nontopological state variables assigned to the nodes.

We ensure that every node in the network is a distinct entity. In
addition, as theoretically the sum of the shareholdings of a company
should be 100\%, we normalize the ownership percentages if the sum is
smaller due to unreported shareholdings. Such missing ownership data
is nearly always due to their percentage values being very small and
hence negligible.

\section{Three-Level Network Analysis}
\label{analysis}

Not all networks can be associated with a notion of flow. For instance,
in the international trade network the fact that country A exports to B
and B exports to C does not imply that goods are flowing from A to C. In
contrast, in ownership networks the distance between two nodes (along a
directed path) corresponds to a precise economic meaning which can be
captured in a measure of control that considers all directed paths of all
lengths (see Sec. \ref{sec:nontop}). In addition, the weight of an ownership
link has a meaning relative to the weight of the other links attached to
the same node. Finally, the value of the nodes themselves is very
important. Therefore, in the following, we focus on network measures
which take these aspects into account, and we do not report on standard
measures such as degree distribution, assortativity, clustering
coefficients, average path lengths, connected components, etc.

\subsection{Level 1: Topological analysis}
\label{sec:top}

\begin{figure}[tH]
\centering
\includegraphics[width=0.25\textwidth]{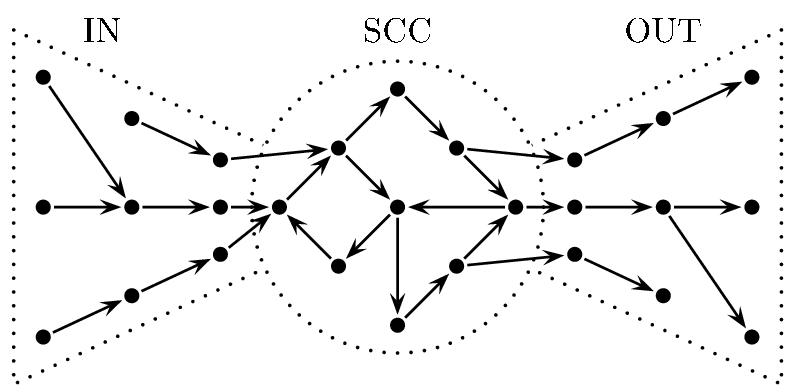}
\caption{Schematic illustration of a bow-tie topology:  the central area is the strongly 
connected component (SCC), where there is a path from each node to every
other node, and the left (IN) and right (OUT) sections contain the incoming and
outgoing nodes, respectively.}\label{fig:bt}
\end{figure}

We start from the analysis of strongly connected components. These
subgraphs correspond to sets of corporations where every firm is
connected to every other firm via a path of indirect ownership.
Furthermore, strongly connected components may form bow-tie structures, akin to the
topology of the world wide web \cite{broder00}. Fig. \ref{fig:bt}
illustrates an idealized bow-tie topology. This structure reflects the
flow of control, as every shareholder in the IN section exerts control
and all corporations in the OUT section are controlled.

We find that roughly two thirds of the countries' ownership networks
contain bow-tie structures (see also \cite{Vitali08}). As an example,
the countries with the highest occurrence of (small) bow-tie
structures are KR and TW, and to a lesser degree JP. A possible
determinant is the well-known existence of so-called business groups
in these countries (e.g., the {\it keiretsu} in JP, and the {\it
chaebol} in KR) forming a tightly knit web of cross shareholdings
\cite{Granovetter95, Feenstra99}. For AU, CA, GB, and US we observe
very few but large bow-tie structures of which the biggest ones
contain hundreds to thousands of corporations. This raises the
question relevant to economics: if the emergence of these
mega-structures in the Anglo-Saxon countries is due to their unique
``type'' of capitalism (the so-called Atlantic or stock market
capitalism \cite{Dore02}), and whether this finding contradicts the
assumption that these markets are characterized by the absence of
business groups \cite{Granovetter95}.

\subsection{Level 2: Extending the notions of degree}
\label{sec:degree}

\begin{figure}[tH]
\centering
\includegraphics[width=0.175\textwidth]{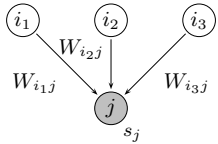}
\caption{Definition of the concentration index $s_j$, measuring the number of
prominent incoming edges, respectively, the effective number of
shareholders of the stock $j$.  When all the weights are equal, then
$s_j=k^{in}_j$, where $k^{in}_j$ is the in degree of vertex $j$. When
one weight is overwhelmingly larger than the others, the concentration
index approaches the value one, meaning that there exists a single
dominant shareholder of $j$. }
\label{fig:Sj}
\end{figure}

In graph theory, the number $k_i$ of edges per vertex $i$ is called
the {\em degree}.  If the edges are oriented, one has to distinguish
between the in degree and out degree, $k^{in}$ and $k^{out}$,
respectively. When the edges $ij$ are weighted with the number
$w_{ij}$, the corresponding quantity is called {\it strength}
\cite{barrat04a}:
\begin{equation}
k^w_i := \sum_j W_{ij}.
\end{equation}

When there are no weights associated with the edges, we expect all
edges to count the same. If weights have a large variance, some edges
will be more important than others. A way of measuring the number of
prominent incoming edges is to define the
\textit{concentration index} \cite{battiston2004isc} as follows:
\begin{equation}
\label{eq:Sj}
s_j := \frac{\left(\sum_{i=1}^{k^{in}_j} W_{ij}\right)^2}{\sum_{i=1}^{k^{in}_j} W_{ij}^2}.
\end{equation}
Note that this quantity is akin to the inverse of the Herfindahl index
extensively used in economics as a standard indicator of market
concentration \cite{Herfindahl59}. Notably, a similar measure has also
been used in statistical physics as an order parameter
\cite{Derrida86}. A recent study \cite{serrano2009emb} employs a
Herfindahl index in their backbone extraction method for weighted
directed networks (where, however, the nodes hold no nontopological
information). In the context of ownership networks, $s_j$ is interpreted
as the effective number of shareholders of the stock $j$, as explained in
Fig. \ref{fig:Sj}. Thus it can be understood as a measure of control
from the point of view of a stock.

\begin{figure}[tH]
\centering
\includegraphics[width=0.29\textwidth]{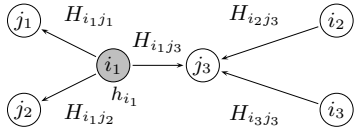}
\caption{The definition of the control index $h_i$, measuring the number of
prominent outgoing edges. In the context of ownership networks this
value represents the effective number of stocks that are controlled by
shareholder $i$. Note that to obtain such a measure, we have to
consider the fraction of control $H_{ij}$, which is a model of how
ownership can be mapped to control (see the discussion in \ref{app:interpret}).}
\label{fig:Hi}
\end{figure}

The second quantity to be introduced measures the number of important
outgoing edges of the vertices. For a given vertex $i$, with a
destination vertex $j$, we first define a measure which reflects the
importance of $i$ with respect to all vertices connecting to $j$:
\begin{equation}
\label{eq:hij}
H_{ij} := \frac{W_{ij}^{2}}{\sum_{l=1}^{k^{in}_j} W_{lj}^{2}}.
\end{equation}
This quantity has values in the interval $(0, 1]$. For instance, if
$H_{ij} \approx 1$ then $i$ is by far the most important source vertex
for the vertex $j$. For our ownership network, $H_{ij}$ represents the
{\it fraction of control} \cite{battiston2004isc} shareholder $i$ has
on the company $j$. As shown in Fig. \ref{fig:Hi}, this quantity is a
way of measuring how important the outgoing edges of a node $i$ are
with respect to its neighbors' neighbors. For an interpretation of
$H_{ij}$ from an economics point of view, consult
\ref{app:interpret}.

From that, we then define the \textit{control index},
\begin{equation}
\label{eq:Hi}
 h_i := \sum_{j=1}^{k^{out}_i} H_{ij}.
\end{equation}
Within the ownership network setting, $h_i$ is interpreted as the
effective number of stocks controlled by shareholder $i$.

\subsection{Distributions of $s$ and $h$}
\label{sec:distributions-S-H}

\begin{figure}[tH]
\centering
\psfrag{Log S}[][][0.65]{$\ln s$}
\psfrag{Log PDF}[][][0.65]{PDF}
\includegraphics[width=0.4\textwidth,angle=-90]{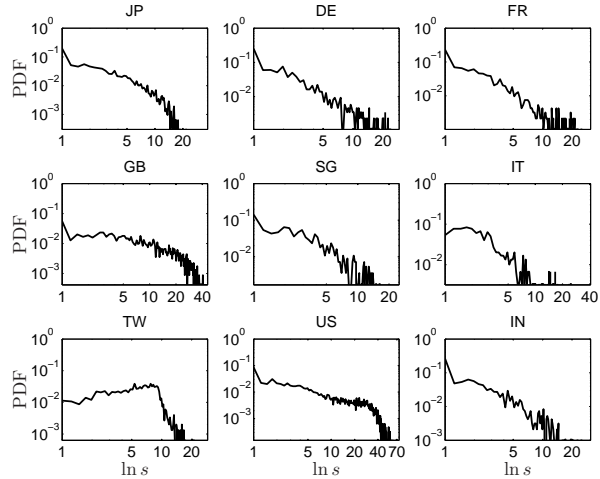}
\caption{Probability distributions of $s_j$ for selected countries; PDF in log-log scale.}\label{fig:S}
\end{figure}

Fig. \ref{fig:S} shows the probability density function (PDF) of $s_j$
for a selection of nine countries (for the full sample consult
\cite{Supp1}). There is a diversity in the shapes and ranges of the distributions to
be seen. For instance, the distribution of GB reveals that many
companies have more than 20 leading shareholders, whereas in IT few
companies are held by more than five significant shareholders. Such
country-specific signatures were expected to appear due to the
differences in legal and institutional settings (e.g., law
enforcement and protection of minority shareholders
\cite{LaPortaSchleifer99}).

\begin{figure}[b!]
\center
\psfrag{h}[][][0.65]{$\ln h$}
\psfrag{kout}[][][0.65]{$\ln k^{out}$}
\psfrag{PDF}[][][0.65]{PDF}
\psfrag{CDF}[][][0.65]{CDF}
  \includegraphics[width=0.4\textwidth,angle=-90]{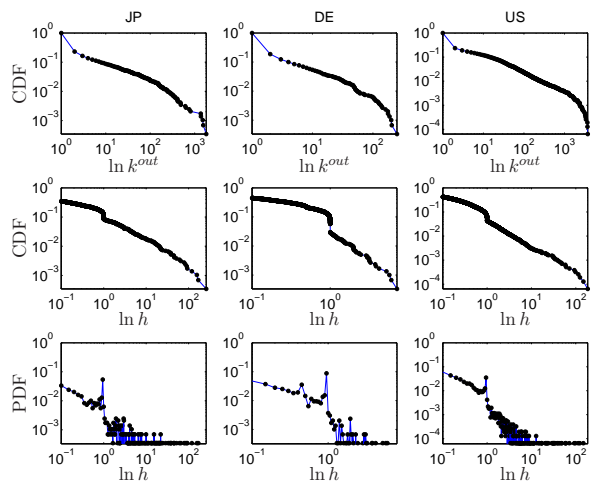}
\caption{Various probability distributions for selected countries: ({\it top panel}) CDF plot of $k_i^{out}$; 
({\it middle panel}) CDF plot of $h_i$; 
 ({\it bottom panel}) PDF plot of $h_i$; all plots are in log-log scale.}\label{fig:H}
\end{figure}

On the other hand, looking at the cumulative distribution function
(CDF) of $k_i^{out}$ (shown for three selected countries
in the top panel of Fig. \ref{fig:H}; the full sample is available at
\cite{Supp1}) a more uniform shape is revealed. The distributions
range across two to three orders of magnitude. Hence some shareholders
can hold up to a couple of thousand stocks, whereas the majority have
ownership in less than 10.  Considering the CDF of $h_i$, seen in the
middle panel of Fig. \ref{fig:H}, one can observe that the curves of
$h_i$ display two regimes. This is true for nearly all analyzed
countries, with a slight country-dependent variability. Notable
exceptions are FI, IS, LU, PT, TN, TW, and VG. In order to understand this
behavior it is useful to look at the PDF of $h_i$, shown in the bottom
panel of Fig. \ref{fig:H}. This uncovers a systematic feature: the
peak at the value of $h_i =1$ indicates that there are many
shareholders in the markets whose only intention is to control one
single stock. This observation, however, could also be due to a
database artifact as incompleteness of the data may result in many
stocks having only one reported shareholder. In order to check that
this result is indeed a feature of the markets, we constrain these
ownership relations to the ones being bigger than 50\%, reflecting
incontestable control. In a subsequent analysis we still observe this
pattern in many countries (BM, CA, CH, DE, FR, GB, ID, IN, KY, MY, TH,
US, and ZA; ES being the most pronounced). In addition, we find many such
shareholders to be non-firms, i.e., people, families, or legal
entities, hardening the evidence for this type of exclusive control.
This result emphasizes the utility of the newly defined measures to
uncover relevant structures in the real-world ownership networks.

\subsection{Level 3: Adding nontopological values}
\label{sec:nontop}

The quantities defined in Eqs. (\ref{eq:Sj}) and (\ref{eq:Hi}) rely on
the direction and weight of the links. However, they do not consider
nontopological state variables assigned to the nodes themselves. In
our case of ownership networks, a natural choice is to use the market
capitalization value of firms in thousand US dollars (USD), $v_j$, as a proxy for
their sizes. Hence $v_j$ will be utilized as the state variable in the
subsequent analysis.  In a first step, we address the question of how
much wealth the shareholders own, i.e., the value in their portfolios.

As the percentage of ownership given by $W_{ij}$ is a measure of the
fraction of outstanding shares $i$ holds in $j$, and the market
capitalization of $j$ is defined by the number of outstanding shares
times the market price, the following quantity reflects $i$'s {\it
portfolio value}:
\begin{equation}
\label{eq:PV}
p_i := \sum_{j=1}^{k^{out}_i} W_{ij} v_j.
\end{equation}
Extending this measure to incorporate the notions of control, we
replace $W_{ij}$ in the previous equation with the fraction of control
$H_{ij}$, defined in Eq. (\ref{eq:hij}), yielding the {\it control
value}:
\begin{equation}
\label{eq:CV}
c_i := \sum_{j=1}^{k^{out}_i} H_{ij} v_j.
\end{equation}
A high $c_i$ value is indicative of the possibility to control a
portfolio with a big market capitalization value. Recall that the
economic meaning of $H_{ij}$ is discussed in \ref{app:interpret}.

It should be noted that Eq. (\ref{eq:CV}) only considers direct
neighbors. To address the question of how control propagates via all
possible direct and indirect ownership paths, the so-called {\it
integrated model} has been proposed \cite{Brioschi89}, which we
briefly sketch.  Consider a sample of $n$ firms connected by
cross-shareholding relations. Let $A_{ij}$, with $i,j=1,2,...,n,$ be
the ownership ($W_{ij}$) or control ($H_{ij}$) that company $i$ has
directly on company $j$, and $A=[A_{ij}]$ is the matrix of all the
links between every one of the $n$ firms. By definition, it holds that
\begin{equation}
  \sum_{i=1}^{n}A_{ij}\leq 1;  \qquad j=1,...,n.
  \label{eq:sum_a}
\end{equation}
When some shareholders of company $i$ are not identified or are outside
the sample $n$, the inequality becomes strict. The integrated model accounts for
direct and indirect  ownership through a recursive computation. The general form of
the equation reads
\begin{equation}
\tilde A_{ij} := A_{ij}+\sum_{n}A_{in}\tilde A_{nj},
\label{eq:int}
\end{equation}
where the tilde denotes integrated ownership or control.
This expression can be written in matrix form as
\begin{equation}
\tilde A=A+A \tilde A,
\label{eq:control}
\end{equation}
the solution of which is given by
\begin{equation}
\tilde A=(I-A)^{-1} A.
\label{eq:control_matrix}
\end{equation}
For the matrix $(I-A)$ to be non-negative and non-singular, a sufficient
condition is that the Frobenius root is smaller than one,
\mbox{$\lambda(A)<1$}. This is ensured by the following requirement: in
each strongly connected component $\mathcal{S}$ there exists at least
one node $j$ such that $\sum_{i\in \mathcal{S}} A_{ij}<1$. In an economic
setting, this means that there exists no subset of $k$ firms $(k =
1,\dots , n)$ that are entirely owned by the $k$ firms themselves. A
condition which is always fulfilled in ownership networks
\cite{Brioschi89}.

In order to define the {\it integrated control value} $\tilde {c}_i$
in the same spirit as Eq. (\ref{eq:CV}), we first solve
Eq. (\ref{eq:control_matrix}) for the fraction of control $H_{ij}$,
which yields the integrated fraction of control $\tilde{H}_{ij}$.
$\tilde {c}_i$ represents the value of control a shareholder gains
from companies reached by all direct and indirect paths of ownership:
\begin{equation}
\label{eq:CVind}
\tilde {c}_i := \sum_{j=1}^{k^{out}_i} \tilde H_{ij} v_j.
\end{equation}
This quantity is used in the next section to identify and rank
the shareholders by importance.

\section{Identifying the Backbone of Corporate Control}
\label{sec:bb}

\subsection{Computing cumulative control}
\label{cumcontr}

The first step of our methodology requires the construction of a
Lorenz-like curve in order to uncover the distribution of the control in
a market. In economics, the Lorenz curve gives a graphical
representation of the cumulative distribution function of a
probability distribution.  It is often used to represent income
distributions, where the $x$ axis ranks the poorest $x \%$ of
households and relates them to a percentage value of income on the
$y$ axis.

Here, on the $x$ axis we rank the shareholders according to their
importance --- as measured by their integrated control value
$\tilde{c}_i$, cf., Eqs. (\ref{eq:hij}), (\ref{eq:control_matrix}), and
(\ref{eq:CVind}) --- and report the fraction they represent with
respect to the whole set of shareholder. The $y$ axis shows the
corresponding percentage of controlled market value, defined as the
fraction of the total market value they collectively or cumulatively
control.

In order to motivate the notion of cumulative control, some preliminary
remarks are required.  Using the integrated control value to rank the
shareholders means that we implicitly assume control based on the
integrated fraction of control $\tilde H_{ij}$. This however is a
potential value reflecting {\it possible} control. In order to
identify the backbone, we take a very conservative approach to the
question of what the {\it actual} control of a shareholder is. To this
aim, we introduce a stringent threshold of 50\%. Any shareholder with
an ownership percentage $W_{ij} > 0.5$ controls by default. This
strict notion of control for a single shareholder is then generalized
to apply to the cumulative control a group of shareholders can
exert. Namely, by requiring the sum of ownership percentages multiple
shareholders have in a common stock to exceed the threshold of
cumulative control. Its value is equivalently chosen to be 50\%.

\begin{figure}[t!]
\center
\includegraphics[width=0.36\textwidth]{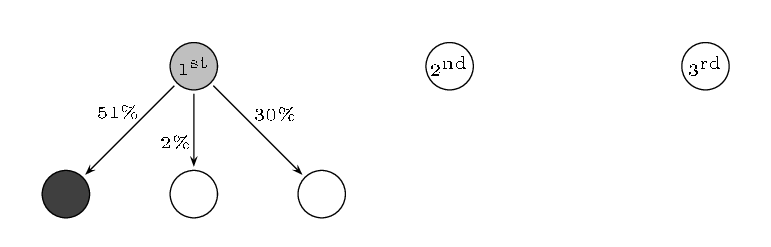}
  \vfill{\phantom{.}}
\includegraphics[width=0.36\textwidth]{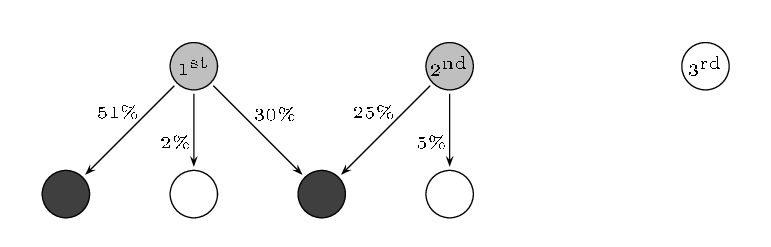}
\caption{
First steps in computing cumulative control: ({\it top panel})
selecting the most important shareholder (light shading) ranked
according to the $\tilde{c}_i$ values and the portfolio of stocks
owned at more than 50\% (dark shading); in the second step ({\it bottom
panel}), the next most important shareholder is added; although there
are now no new stocks which are owned directly at more than 50\%,
cumulatively the two shareholder own an additional stock at 55\%.
}\label{fig:ccalg}
\end{figure}

We start the computation of cumulative control by identifying the
shareholder  having the highest $\tilde{c}_i$ value. From the
portfolio of this holder, we extract the stocks that are owned at more
than the said 50\%. In the next step, the shareholder with the second
highest $\tilde{c}_i$ value is selected. Next to the stocks
individually held at more than 50\% by this shareholder, additional
stocks are considered, which are cumulatively owned by the top two
shareholders at more than the said threshold value. See
Fig. \ref{fig:ccalg} for an illustrated example.

$U_{in} (n)$ is defined to be the set of indices of the stocks that
are individually held above the threshold value by the $n$ selected
top shareholders.  Equivalently, $U_{cu}(n)$ represents the set of
indices of the cumulatively controlled companies. It holds that $U_{in}
(n) \cap U_{cu} (n) = \emptyset$. At each step $n$, the total value of
this newly constructed portfolio, $U_{in} (n) \cup U_{cu} (n)$, is
computed:
\begin{equation}
\label{eq:vn}
v_{cu} (n) := \sum_{j \in  U_{in}(n)}  v_j + \sum_{j \in U_{cu}(n)}v_j.
\end{equation}
Eq. (\ref{eq:vn}) is in contrast to Eq. (\ref{eq:PV}), where the total
value of the stocks $j$ is multiplied by the ownership percentage
$W_{ij}$.

Let $n_{tot}$ be the total number of shareholders in a market and
$v_{tot}$ the total market value.  We normalize with these values,
defining:
\begin{equation}
\eta(n) := \frac{n}{n_{tot}} , \qquad \vartheta(n) := \frac{v_{cu}(n)}{v_{tot}},
\end{equation}
where $\eta,\vartheta \in (0,1]$.

\begin{figure}[t!]
  \psfrag{YLAB}[][][0.75]{$\vartheta$ (Market Value, \%)}
  \psfrag{XLAB}[][][0.75]{$\eta$ (Shareholder Rank, \%)}
  \psfrag{TITLE}[b][][1.0]{Cumulative Control} \centering
\includegraphics[width=0.4\textwidth,angle=-90]{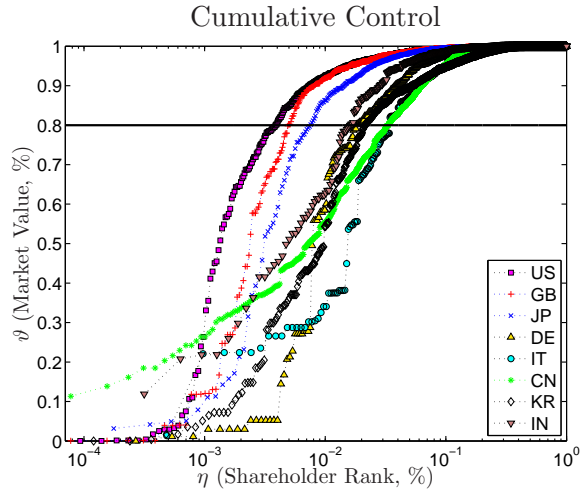}
\caption{(Color online) Fraction of shareholders $\eta$, sorted by
  descending (integrated) control value $\tilde{c}_i$,
  cumulatively controlling $\vartheta$ percent of the total market value; the
  horizontal line denotes a market value of 80\%; the diagram is in
  semilogarithmic scale.}\label{fig:cc}
\end{figure}

In Fig. (\ref{fig:cc}) these values are plotted against each other for
a selection of countries (the full sample is in \cite{Supp1}),
yielding the cumulative control diagram, akin to a Lorenz curve (with
reversed $x$ axis). As an example, a coordinate pair with value
$(10^{-3}, 0.2)$ reveals that the top $0.1\%$ of shareholders
cumulatively control $20\%$ of the total market value. The top right
corner of the diagram represents 100\% of the shareholders
controlling 100\% of the market value, and the first data point in the
lower left-hand corner denotes the most important shareholder of each
country. Different countries show a varying degree of concentration of
control.

It should be emphasized that our analysis unveils the importance of
shareholders: the ranking of every shareholder is based on all direct
and indirect paths of control of any length. In contrast, most other
empirical studies start their analysis from a set of important stocks
(e.g., ranked by market capitalization). The methods of accounting for
indirect control (see Sec. \ref{sec:nontop}) are, if at all, only
employed to detect the so-called ultimate owners of the stocks. For
instance, \cite{LaPorta98} studies the ten largest corporations in 49
countries, \cite{LaPortaSchleifer99} looks at the 20 largest public
companies in 27 countries, \cite{Claessens00} analyzes 2980 companies
in nine East Asian countries, and \cite{Chapelle05} utilizes a set of
800 Belgian firms.

Finally, note that although the identity of the individual controlling
shareholders is lost due to the introduction of cumulative control,
the emphasis lies on the fact that the controlling shareholders are
present in the set of the first $n$ holders.

\subsection{Extracting the backbone}
\label{extract}

Once the curve of the cumulative control is known for a market, one
can set a threshold for the percentage of jointly controlled market
value, $\hat \vartheta$. This results in the identification of the
percentage $\hat \eta$ of shareholders that theoretically hold the
power to control this value, if they were to coordinate their
activities in corresponding voting blocks.  The subnetwork of these
power holders and their portfolios is called the backbone. Here we
choose the value $\hat \vartheta = 0.8$, revealing the power holders
able to control 80\% of the total market value.

Algorithm (\ref{alg:cc}) gives the complete recipe for computing the
backbone. As inputs, the algorithm requires all the
$\tilde{c}_i$ values, the threshold defining the level of (cumulative)
control $\delta$ and the threshold for the considered market value
$\hat \vartheta$.  Steps 1 -- 7 are required for the cumulative
control computation and $\delta$ is set to $0.5$. Step 8 specifies the
interruption requirement given by the controlled portfolio value being
bigger than $\hat \vartheta$ times the total market value.

\begin{algorithm}[t]
  \caption{$\mathcal{BB}$($\tilde{c}_1, \dots ,\tilde{c}_n$, $\delta$, $\hat \vartheta$ )}
  \begin{algorithmic}[1]
   \State $\tilde{c} \gets sort\_descending(\tilde{c}_1, \dots ,\tilde{c}_n)$ \label{alg:cc}
    \Repeat
                        \State $c \gets get\_largest(\tilde{c})$
                        \State $I \gets I \cup index(c)$
                        \State $PF \gets  stocks\_controlled\_by(I)$ (individually and cumulatively at more than $\delta$)
                        \State $PFV \gets value\_of\_portfolio(PF)$ 
                        \State $\tilde{c} \gets \tilde{c} \setminus \{c\}$
    \Until $PFV \ge \hat \vartheta \cdot total\_market\_value$
  \State $prune\_network(I,PF)$ 
  \end{algorithmic}
\end{algorithm}

Finally, in step 9, the subnetwork of power holders and their new
portfolios is pruned to eliminate weak links and further enhance the
important structures. For each stock $j$ in the union of these
portfolios, only as many shareholders are kept as the rounded value of
$s_j$ indicates, i.e., the (approximate) effective number of
shareholders. Although a power holder can be in the portfolio of other
power holders, the pruning only considers the incoming links.  That is,
if $j$ has five holders but $s_j$ is roughly three, only the three
largest shareholders are considered for the backbone. The portfolio of
$j$ is left untouched. In effect, the weakest links and any resulting
isolated nodes are removed.

\subsection{Generalizing the method of backbone extraction}
\label{sec:generalizing-backbone-method}
Notice that our method can be generalized to any directed and weighted
network in which (1) a nontopological real value $v_j \ge 0$ can be
assigned to the nodes (with the condition that $v_j>0$ for at least
all the leaf nodes in the network) and (2) an edge from node $i$ to
$j$ with weight $W_{ij}$ implies that some of the value of $j$ is
transferred to $i$.  Assume that the nodes which are associated with a
value $v_j$ produce $v_j$ units of mass at time $t=1$. Then the flow
$\phi_i$ entering the node $i$ from each node $j$ at time $t$ is the
fraction $W_{ij}$ of the mass produced directly by $j$ plus the same
fraction of the inflow of $j$:
\begin{equation}
  \label{eq:mass-inflow}
  \phi_i(t+1) = \sum_j W_{ij} v_j + \sum_j W_{ij}  \phi_i(t),
\end{equation}
where $\sum_iW_{ij}=1$ for the nodes $j$ that have predecessors and
$\sum_iW_{ij}=0$ for the root nodes (sinks). In matrix notation, at
the steady state, this yields
\begin{equation}
  \label{eq:mass-inflow-matrix}
  \phi= W  (v + \phi).
\end{equation}
The solution
\begin{equation}
  \label{eq:mass-inflow-solution}
  \phi = (1-W)^{-1} W v,
\end{equation}
exists and is unique if $\lambda (W) <1$. This condition is easily
fulfilled in real networks as it requires that in each strongly
connected component $\mathcal{S}$ there exists at least one node $j$
such that $\sum_{i\in \mathcal{S}} W_{ij}<1$. Or, equivalently, the
mass circulating in $\mathcal{S}$ is also flowing to some node outside
of $\mathcal{S}$. To summarize, some of the nodes only produce mass
(all the leaf nodes but possibly also other nodes) at time $t=1$ and
are thus sources, while the root nodes accumulate the mass. Notice
that the mass is conserved at all nodes except at the sinks.

The convention used in this paper implies that mass flows against the
direction of the edges. This makes sense in the case of ownership
because although the cash allowing an equity stake in a firm to be
held flows in the direction of the edges, control is transferred in
the opposite direction, from the corporation to its shareholders. This
is also true for the paid dividends. Observe that the integrated
control value defined in Eq. (\ref{eq:CVind}) can be written in matrix
notation as
\begin{equation}
  \label{eq:CVind-matrix}
  \tilde c = \tilde H v =  (1-H)^{-1} H v,
\end{equation}
which is in fact equivalent to Eq. (\ref{eq:mass-inflow-solution}). This
implies that for any node $i$ the integrated control value $\tilde c_i =
\sum_j \tilde H_{ij} v_j$ corresponds to the inflow $\phi_i$ of mass in
the steady state. 

Returning to the generic setting, let $U_0$ and $E_0$ be,
respectively, the set of vertices and edges yielding the network. We
define a subset $U \subseteq U_0$ of vertices on which we want to
focus on (in the analysis presented earlier $U=U_0$). Let $E \subseteq
E_0$ then be the set of edges among the vertices in $U$ and introduce
$\hat \vartheta$, a threshold for the fraction of aggregate flow through the
nodes of the network. If the relative importance of neighboring nodes
is crucial, $H_{ij}$ is computed from $W_{ij}$ by the virtue of
Eq. (\ref{eq:hij}). Note that $H_{ij}$ can be replaced by any function
of the weights $W_{ij}$ that is suitable in the context of the network
under examination.  We now solve Eq. (\ref{eq:control_matrix}) to
obtain the integrated value $\tilde{H}_{ij}$. This yields the
quantitative relation of the indirect connections among the
nodes. To be precise, it should be noted that in some networks the
weight of an indirect connection is not correctly captured by the
product of the weights along the path between the two nodes. In such
cases one has to modify Eq. (\ref{eq:int}) accordingly.

The next step in the backbone extraction procedure is to identify the
fraction of flow that is transferred by a subset of nodes. A systematic
way of doing this was presented in Sec. \ref{cumcontr} where we
constructed the curve, $(\eta, \vartheta)$. A general recipe for such
a construction is the following. On the $x$ axis all the nodes are
ranked by their $\phi_i$ value in descending order and the fraction
they represent with respect to size of $U$ is captured.  The $y$ axis
then shows the corresponding percentage of flow the nodes transfer. As
an example, the first $k$ (ranked) nodes represent the fraction
$\eta(k) = k / |U|$ of all nodes that cumulatively transfer the amount
$\vartheta (k) = (\sum_{i=1}^k \phi_i ) / \phi_{tot}$ of the total
flow.  Furthermore, $\hat{\eta}$ corresponds to the percentage of top
ranked nodes that pipe the predefined fraction $\hat \vartheta$ of all
the mass flowing in the whole network. Note that the procedure
described in Sec. \ref{cumcontr} is somewhat different. There we
considered the fraction of the total value given by the direct
successors of the nodes with largest $\tilde{c}_i$. This makes sense
due to the special nature of the ownership networks under
investigation, where every non-firm shareholder (root node) is
directly linked to at least one corporation (leaf node), and the
corporations are connected among themselves.

Consider the union of the nodes identified by $\hat \eta$ and their
direct and indirect successors, together with the links among
them. This is a subnetwork $\mathcal{B}=(U^B,E^B)$, with $U^B\subset
U$ and $E^B\subset E$ that comprises, by construction, the fraction
$\hat \vartheta$ of the total flow. This is a first possible
definition of the backbone of $(U,E)$. A discussion of the potential
application of this procedure to other domains, and a more detailed
description of the generalized methodology (along with specific
refinements pertaining to the context given by the networks) is left
for future work. Viable candidates are the world trade web
\cite{serrano2003twt,GL04b,reichardt2007rmc,fagiolo2008tpw}, food webs
\cite{garlaschelli2003usr}, transportation networks
\cite{kuhnert2006slu}, and credit networks \cite{battiston2007cca}.

\subsection{Defining classification measures}
\label{class}

\begin{figure}[t!]
\centering
\includegraphics[width=0.375\textwidth]{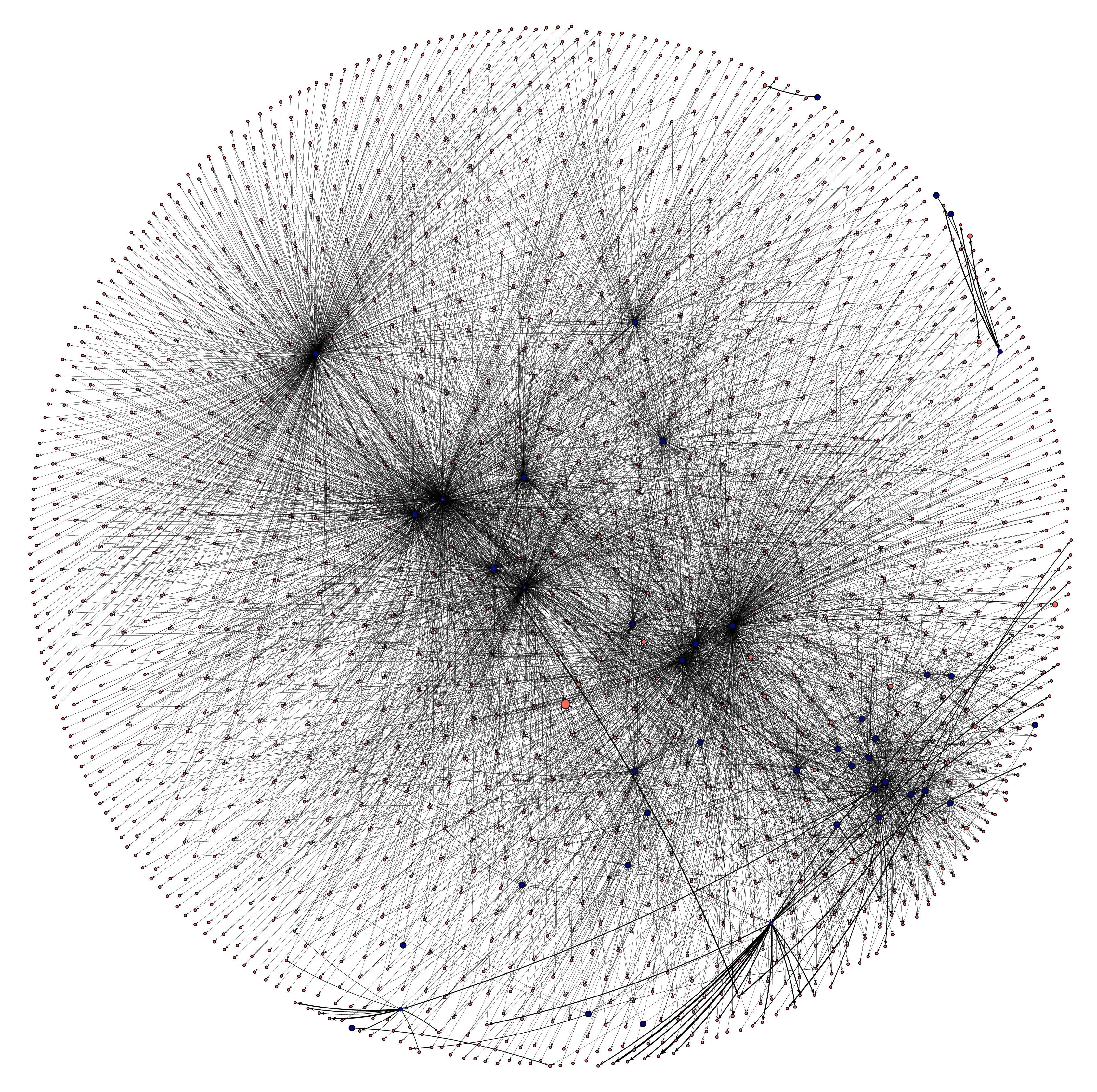}
\vfill{\phantom{.}}
\includegraphics[width=0.375\textwidth]{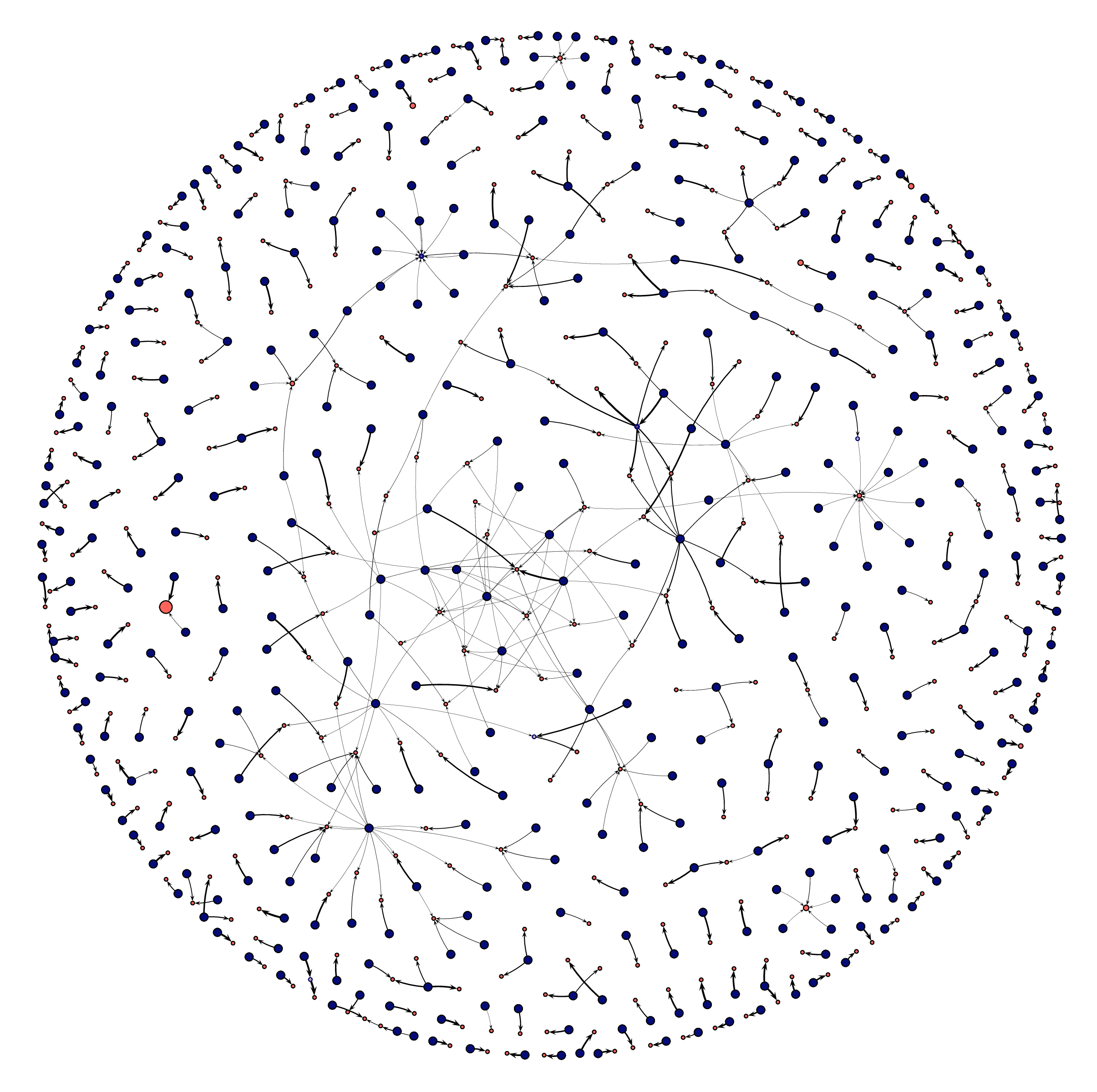}
\caption{({\it Top}) the backbone of JP; ({\it bottom}) the backbone of
  CN (for the complete set of backbone layouts consult \cite{Supp1}); the
  graph layouts are based on \cite{geipel07}.} \label{fig:bb}
 \end{figure}

 According to economists, markets differ from one country to another in a
 variety of respects \cite{LaPorta98, LaPortaSchleifer99}. They may not
 look too different if one restricts the analysis to the distribution of
 local quantities, and in particular to the degree, as shown in
 Sec. \ref{sec:distributions-S-H}. In contrast, at the level of the
 backbones, i.e., the structures where most of the value resides, they
 can look strikingly dissimilar. As seen for instance in the case of CN
 and JP, shown in Fig. \ref{fig:bb}. In the following, we provide a
 quantitative classification of these diverse structures based on the
 indicators used to construct the backbones.

Let $n_{st}$ and $n_{sh}$ denote the number of
stocks and shareholders in a backbone, respectively. As $s_j$ measures
the effective number of shareholders of a company, the average value,
\begin{equation}
\label{eq:sb}
\overline s = \frac{\sum_{j=1}^{n_{st}} s_j}{n_{st}},
\end{equation}
is a good proxy characterizing the local patterns of ownership: the
higher $\overline s$, the more dispersed the ownership is in the
backbone or the more common is the appearance of widely held
firms. Furthermore, due to the construction of $s_j$, the metric
$\overline s$ equivalently measures the local concentration of control.

\begin{figure}[t!]
\center
\includegraphics[width=0.35\textwidth]{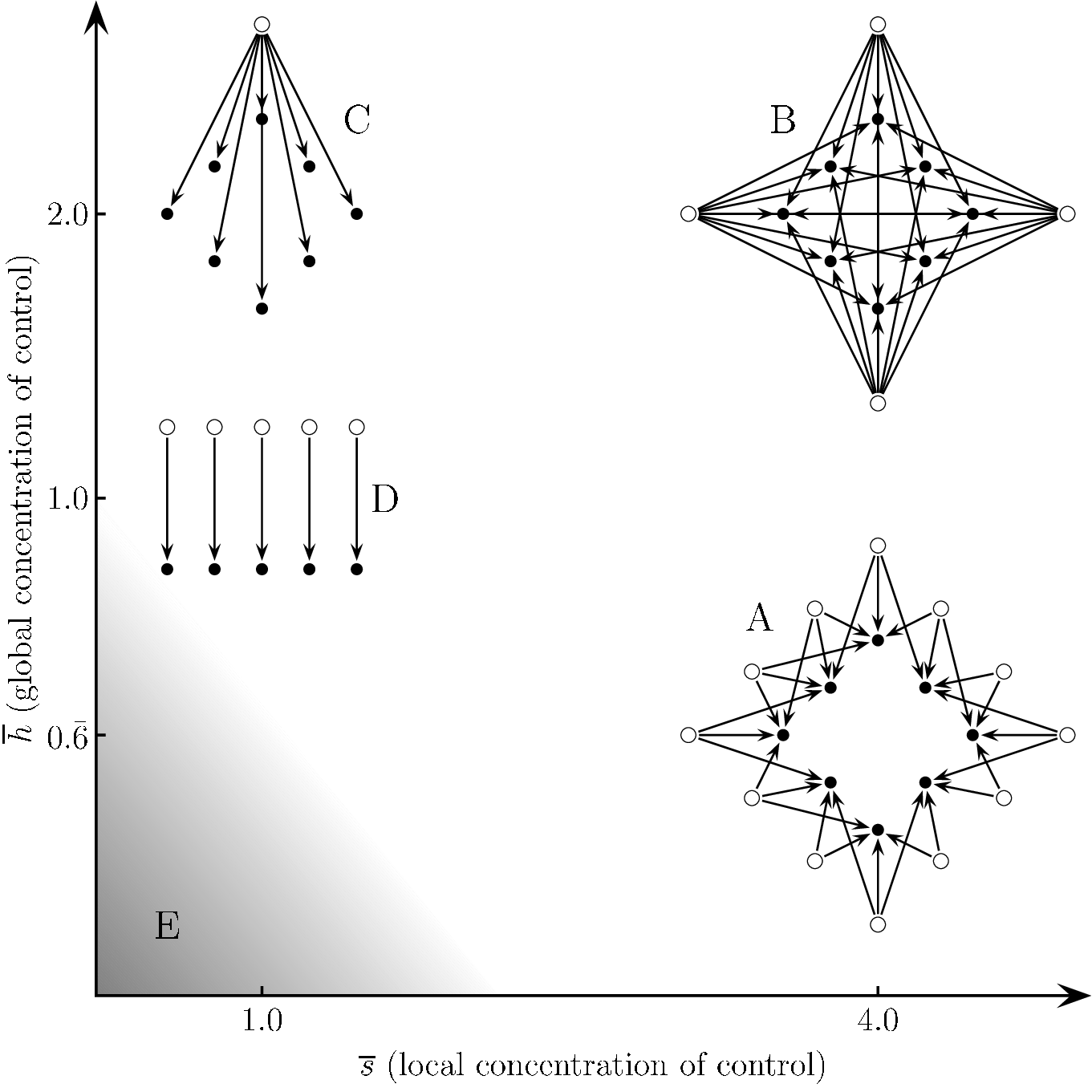}
\caption{The map of control:  illustration of idealized network 
topologies in terms of local dispersion of control ($\overline s$)
vs global concentration of control ($\overline h$); shareholders and
stocks are shown as empty and filled bullets, respectively; arrows
represent ownership; region (E) is excluded due to consistency
constraints; (A) does not necessarily need to be a single connected
structure; see Fig. \ref{fig:sh} for the empirical results.
}\label{fig:map}
\end{figure}

In a similar vein, the average value
\begin{equation}
\label{eq:hb}
\overline h = \frac{\sum_{i=1}^{n_{sh}} h_i}{n_{sh}} = \frac{n_{st}}{n_{sh}},
\end{equation}
reflects the global distribution of control. A high value of $\overline
h$ means that the considered backbone has very few shareholders compared
to stocks, exposing a high degree of global concentration of
control. Recall that $n_{st}$ and $n_{sh}$ refer to the backbone and not
to the original network. Fig. \ref{fig:map} shows the possible generic
backbone configurations resulting from local and global distributions of
control.

Remember also that in order to construct the backbones we had to specify
a threshold for the controlled market value: $\hat \vartheta = 0.8$. In
the cumulative control diagram seen in Fig. (\ref{fig:cc}), this allows
the identification of the number of shareholders being able to control
this value. The value $\hat \eta$ reflects the percentage of
power holders corresponding to $\hat \vartheta$. To adjust for the
variability introduced by the different numbers of shareholders present
in the various national stock markets, we chose to normalize $\hat
\eta$. Let $n_{100}$ denote the smallest number of shareholders
controlling 100\% of the total market value $v_{tot}$, then
\begin{equation}
\label{eq:hhn}
\eta^\prime := \frac{\hat \eta}{n_{100}}.
\end{equation}
A small value for $\eta^\prime$ means that there will be very few
shareholders in the backbone compared to the number of shareholders
present in the whole market, reflecting that the market value is
extremely concentrated in the hands of a few shareholders. In essence,
the metric $\eta^\prime$ is an emergent property of the backbone extraction
algorithm and mirrors the global distribution of the value.

\section{Analyzing the Backbones}
\label{bbanalysis}

\begin{figure}[tH]
\centering
\includegraphics[width=0.32\textwidth]{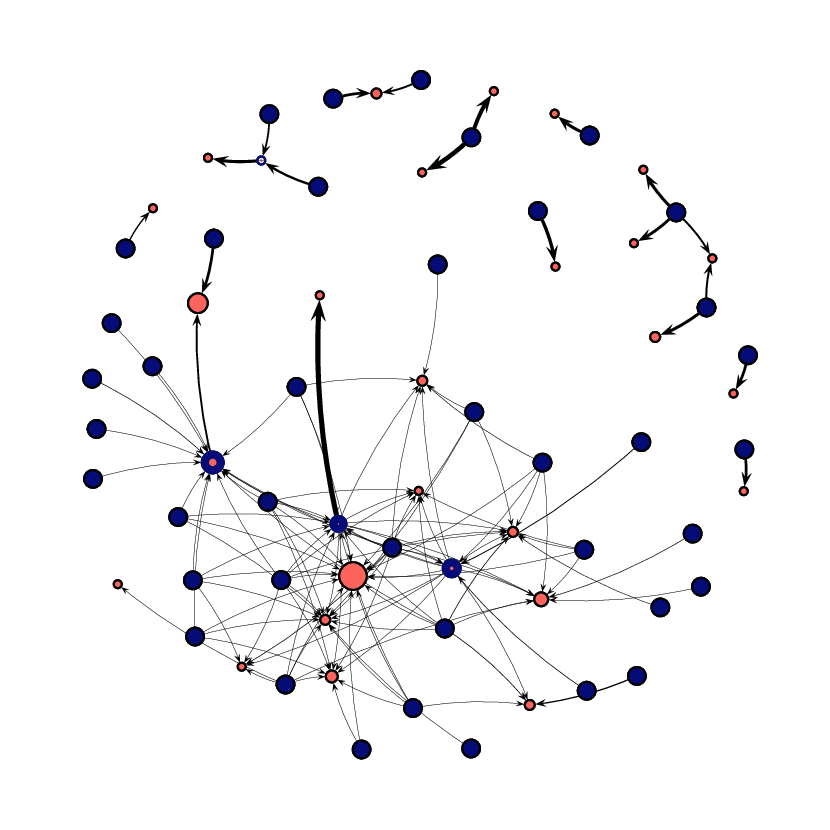}
\caption{ 
(Color online) The backbone of CH is a subnetwork of the original
ownership network which was comprised of 972 shareholders, 266 stocks,
and 4671 ownership relations; firms are denoted by red nodes and sized
by market capitalization, shareholders are blue, whereas firms owning
stocks themselves are represented by red nodes with thick blue
bounding circles, arrows are weighted by the percentage of ownership
value; the graph layouts are based on \cite{geipel07}.} \label{fig:ch}
\end{figure}

How relevant are the backbones and how many properties of the
real-world ownership networks are captured by the classification
measure? As a qualitative example, Fig. (\ref{fig:ch}) shows the
layout for the CH backbone network. Looking at the few stocks left in
the backbone, it is indeed the case that the important corporations
reappear (recall that the algorithm selected the shareholders).  We
find a cluster of shareholdings linking, for instance, Nestl\'e,
Novartis, Roche Holding, UBS, Credit Suisse Group, ABB, Swiss Re,
and Swatch. JPMorgan Chase \& Co. features as the most important controlling
shareholder.  The descendants of the founding families of Roche
(Hoffmann and Oeri) are the highest ranked Swiss shareholders at
position four. UBS follows as dominant Swiss shareholder at rank
seven.

We can also recover some previous empirical results. The ``widely held''
index \cite{LaPortaSchleifer99} assigns to a country a value of one if
there are no controlling shareholders, and zero if all firms in the
sample are controlled above a given threshold. The study is done with a
$10\%$ and $20\%$ cut-off value for the threshold.  We find a $76.6\%$
correlation (and a $p$ value for testing the hypothesis of no correlation
of $3.2 \times 10^{-6}$) between $\overline s$ in the backbones and the
$10\%$ cut-off ``widely held'' index for the 27 countries it is reported
for. The correlation of $\overline s$ in the countries' whole ownership
networks is $60.0\%$ ($9.3 \times 10^{-4}$). For the $20\%$ cutoff, the
correlation values are smaller. These relations should however be handled
with care, as the study \cite{LaPortaSchleifer99} is restricted to the 20
largest firms (in terms of market capitalization) in the analyzed
countries and there is a 12 year lag between the data sets in the two
studies. 

The backbone extraction algorithm is also a good test for the robustness
of market patterns. The bow-tie structures (discussed in
Sec. \ref{sec:top}) in JP, KR, and TW vanish or are negligibly small in their
backbones, whereas in the backbones of the Anglo-Saxon countries (and as
an outlier SE) one sizable bow-tie structure survives. This emphasizes
the strength and hence the importance of these patterns in the markets of
AU, CA, GB, and US.

\subsection{Global concentration of control}
\label{global}

\begin{figure}[tH]
\center
\psfrag{log(S)}{$\ln(\overline s)$}
\psfrag{log(H)}{$\ln(\overline h)$}
           \includegraphics[width=0.38\textwidth,angle=-90]{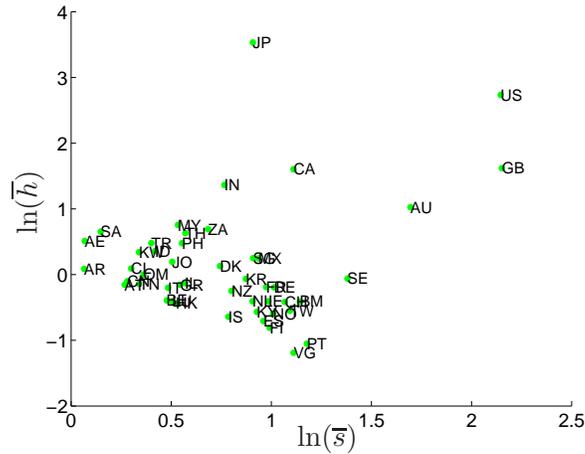}
\caption{Map of control: local dispersion of control, $\overline s$, is 
plotted against global concentration of control, $\overline h$, for 48 countries.}
\label{fig:sh}
\end{figure}

We utilize the measures defined in Sec. \ref{class} to classify the 48
backbones. In Fig. \ref{fig:sh} the logarithmic values of $\overline s$ and
$\overline h$ are plotted against each other. $\overline s$ is a local
measure for the dispersion of control (at first-neighbor level,
insensitive to value).  A large value indicates a high presence of widely
held firms. $\overline h$ is an indicator of the global concentration of
control [an integrated measure, i.e., derived by virtue of
Eq. (\ref{eq:control_matrix}), at second-neighbor level, insensitive to
value]. Large values are indicative that the control of many stocks
resides in the hands of very few shareholders. The $\overline
s$ coordinates of the countries are as expected
\cite{LaPortaSchleifer99}: to the right we see countries known to have
widely held firms (AU, GB, and US). Instead, FR, IT, and JP are located to
the left, reflecting more concentrated local control. However, there is a
counterintuitive trend in the data: the more local control is dispersed,
the higher the global concentration of control becomes. What looks like a
democratic distribution of control from close up, actually turns out to
warp into highly concentrated control in the hands of very few
shareholders. On the other hand, the local concentration of control is in
fact widely distributed among many controlling shareholder. Comparing
with Fig. \ref{fig:map}, where idealized network configurations are
illustrated, we conclude that the empirical patterns of local and global
control correspond to network topologies ranging from types (B) to type
(D), with JP combining local and global concentration of
control. Interestingly, type (A) and (C) constellations are not observed
in the data.

\begin{figure}[tH]
\psfrag{log(S)}{$\ln(\overline s)$}
\psfrag{log(hhat)}{$\ln(\eta^\prime)$}
\center
           \includegraphics[width=0.38\textwidth, angle=-90]{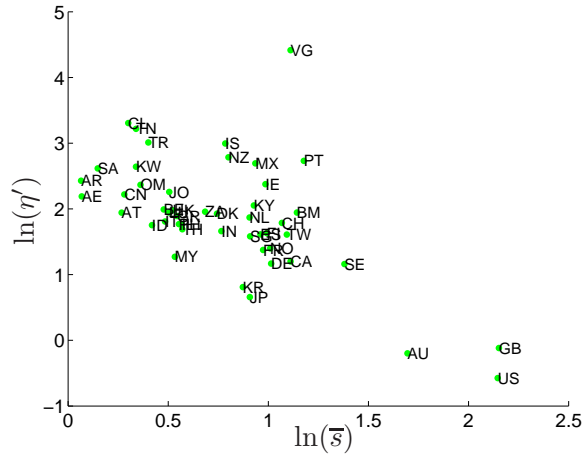}
\caption{Map of market value: local dispersion of control, $\overline s$, is 
plotted against global concentration of market value, $\eta^\prime$, for 48 countries.}
\label{fig:shh}
\end{figure}

In Fig. \ref{fig:shh} the logarithmic values of $\overline s$ and $\eta^\prime$
are depicted. $\eta^\prime$ is a global variable related to the
(normalized) percentage of shareholders in the backbone (an emergent
quantity). It hence measures the concentration of value in a market, as a
low number means that very few shareholders are able to control $80\%$ of
the market value. What we concluded in the last paragraph for control is
also true for the market value: the more the control is locally dispersed,
the higher the concentration of value that lies in the hands of very few
controlling shareholders and vice versa. 

We realize that the two figures discussed in this section open many
questions. Why are there outliers such as JP in Fig. \ref{fig:sh} and VG
in Fig. \ref{fig:shh}? What does it mean to group countries according to
their $\overline s$, $\overline h$, and $\eta^\prime$ coordinates and what
does proximity imply? What are the implications for the individual
countries? We hope to address such and similar questions in future work.

\subsection{Seat of power}
\label{power}

Having identified important shareholders in the global markets, it
is now also possible to address the following questions.  Who holds
the power in an increasingly globalized world? How important are
individual people compared to the sphere of influence of multinational
corporations? How eminent is the influence of the financial sector? By
looking in detail at the identity of the power holders featured in the
backbones, we address these issues next.

If one focuses on how often the same power holders appear in the
backbones of the 48 countries analyzed, it is possible to identify the
global power holders. Following is a top-ten list, comprised of the
company's name, activity, country the headquarter is based in, and ranked
according to the number of times it is present in different countries'
backbones: the Capital Group Companies (investment management, US, 36),
Fidelity Management \& Research (investment products and services, US,
32), Barclays PLC (financial services provider, GB, 26), Franklin
Resources (investment management, US, 25), AXA (insurance company, FR,
22), JPMorgan Chase \& Co. (financial services provider, US, 19),
Dimensional Fund Advisors (investment management, US, 15), Merrill Lynch
\& Co. (investment management, US, 14), Wellington Management Co.
(investment management, US, 14), and UBS (financial services provider, CH,
12). 

Next to the dominance of US American companies we find: Barclays PLC
(GB), AXA (FR) and UBS (CH), Deutsche Bank (DE), Brandes Investment
Partners (CA), Soci\'et\'e G\'en\'erale (FR), Credit Suisse Group (CH),
Schroders PLC (GB), and Allianz (DE) in the top 21 positions.  The government
of Singapore is at rank 25. HSBC Holdings PLC (HK/GB), the world's
largest banking group, only appears at position 26. In addition, large
multinational corporations outside of the finance and insurance industry
do not act as prominent shareholders and only appear in their own
national countries' backbones as controlled stocks. For instance, Exxon
Mobil, Daimler Chrysler, Ford Motor Co., Siemens, and Unilever.

Individual people do not appear as multinational power holders very
often. In the US backbone, we find one person ranked at ninth position:
Warren E. Buffet. William Henry Gates III is next, at rank 26.  In DE the
family Porsche/Piech and in FR the family Bettencourt are power-holders
in the top ten. For the tax-haven KY one finds Kao H. Min (who is placed
at number 140 in the Forbes 400 list) in the top ranks.

The prevalence of multinational financial corporations in the list above
is perhaps not very surprising. For instance, Capital Group Companies is
one of the world's largest investment management organizations with
assets under management in excess of one trillion USD. However, it is an
interesting and novel observation that all the above-mentioned
corporations appear as prominent {\it controlling} shareholders
simultaneously in many countries. We are aware that financial
institutions such as mutual funds may not always seek to exert overt
control. This is argued, for instance, for some of the largest US mutual
funds when operating in the US \cite{davis2008nfc}, on the basis of their
propensity to vote against the management (although, the same mutual
funds are described as exerting their power when operating in
Europe). However, to our knowledge, there are no systematic studies about
the control of financial institutions over their owned companies
world wide. To conclude, one can interpret our quantitative measure of
control as potential power (namely, the probability of achieving one's
own interest against the opposition of other actors
\cite{weber1997tsa}). Given these premises, we cannot exclude that the
top shareholders having vast potential power do not globally exert
it in some way.

\section{Summary and Conclusion}
\label{conc}

We have developed a methodology to identify and extract the backbone
of complex networks that are comprised of weighted and directed links
and nodes to which a scalar quantity is associated. We interpret such
networks as systems in which mass is created at some nodes and
transferred to the nodes upstream. The amount of mass flowing along a
link from node $i$ to node $j$ is given by the scalar quantity
associated with the node $j$ times the weight of the link, $W_{ij} \,
v_j$. The backbone corresponds to the subnetwork in which a
preassigned fraction of the total flow of the system is transferred.

Applied to ownership networks, the procedure identifies the backbone
as the subnetwork where most of the control and the economic value
resides. In the analysis the nodes are associated with nontopological
state variables given by the market capitalization value of the firms,
and the indirect control along all ownership pathways is fully
accounted for.  We ranked the shareholders according to the value they
can control, and we constructed the subset of shareholders which
collectively control a given fraction of the economic value in the
market. In essence, our algorithm for extracting the backbone
amplifies subtle effects and unveils key structures. We further
introduced some measures aimed at classifying the backbone of the
different markets in terms of local and global concentration of
control and value. We find that each level of detail in the analysis
uncovers features in the ownership networks. Incorporating the
direction of links in the study reveals bow-tie structures in the
network. Including value allows us to identify who is holding the
power in the global stock markets.

With respect to other studies in the economics literature, next to
proposing a model for estimating control from ownership, we are able
to recover previously observed patterns in the data, namely, the frequency
of widely held firms in the various countries studied. Indeed, it has
been known for over 75 years that the Anglo-Saxon countries have the
highest occurrence of widely held firms \cite{BerleMeans32}. The
statement that the control of corporations is dispersed among many
shareholders invokes the intuition that there exists a multitude of
owners that only hold a small amount of shares in a few
companies. However, in contrast to such intuition, our main finding is
that a local dispersion of control is associated with a global
concentration of control and value. This means that only a small elite of
shareholders controls a large fraction of the stock market, without ever
having been previously systematically reported on. Some authors have
suggested such a result by observing that a few big US mutual funds
managing personal pension plans have become the biggest owners of
corporate America since the 1990s \cite{davis2008nfc}. On the other hand,
in countries with local concentration of control (mostly observed in
European states), the shareholders tend to only hold control over a
single corporation, resulting in the dispersion of global control and
value.  Finally, we also observe that the US financial sector holds the
seat of power at an international level. It will remain to be seen, if
the continued unfolding of the current financial crisis will tip this
balance of power as the US financial landscape faces a fundamental
transformation in its wake.


\section*{Acknowledgements}
We would like to express our special gratitude to G. Caldarelli and
D. Garlaschelli who provided invaluable advice to this research
especially in its early stages. We would also like to thank F. Schweitzer
and M. Napoletano for fruitful discussions. Finally, we are very grateful
for the advice of G. Davis regarding the relevance of our work with
respect to issues in corporate governance.

\begin{appendix}

\renewcommand\thesection{Appendix \Alph{section}}

\section{Analyzed Countries}
\label{app:countries}

Data from the following countries was used: United Arab Emirates (AE),
Argentina (AR), Austria (AT), Australia (AU), Belgium (BE), Bermuda
(BM), Canada (CA), Switzerland (CH), Chile (CL), China (CN), Germany
(DE), Denmark (DK), Spain (ES), Finland (FI), France (FR), United
Kingdom (GB), Greece (GR), Hong Kong (HK), Indonesia (ID), Ireland
(IE), Israel (IL), India (IN), Iceland (IS), Italy (IT), Jordan (JO),
Japan (JP), South Korea (KR), Kuwait (KW), Cayman Islands (KY),
Luxembourg (LU), Mexico (MX), Malaysia (MY), Netherlands (NL), Norway
(NO), New Zealand (NZ), Oman (OM), Philippines (PH), Portugal (PT),
Saudi Arabia (SA), Sweden (SE), Singapore (SG), Thailand (TH), Tunisia
(TN), Turkey (TR), Taiwan (TW), USA (US), Virgin Islands (VG), and South
Africa (ZA).

Countries are identified by their two letter ISO 3166--1 alpha-2 codes
(given in the parenthesis above).

\section{Ownership vs. Control or the Interpretation of $H_{ij}$}
\label{app:interpret}

While ownership is an objective quantity (the percentage of shares
owned), control (reflected in voting rights) can only be estimated.
In this appendix we provide a motivation for our proposed model of
control $H_{ij}$ (defined in Sec. \ref{sec:degree}) from an economics
point of view and discuss how our measure overcomes some of the
limitations of previous models.

There is a great freedom in how corporations are allowed to map
percentages of ownership in their equity capital (also referred to as
cash-flow rights) into voting rights assigned to the holders at
shareholders meetings. However, empirical studies indicate that in
many countries the corporations tend not to exploit all the
opportunities allowed by national laws to skew voting rights. Instead,
they adopt the so-called one-share-one-vote principle which states
that ownership percentages yield identical percentages of voting
rights \cite{LaPortaSchleifer99,deminor05}.

It is however still not obvious how to compute control from the
knowledge of the voting rights. As an example, some simple models
introducing a fixed threshold for control have been proposed (with
threshold values of 10\% and 20\% \cite{LaPortaSchleifer99} next to a
more conservative value of 50\% \cite{Chapelle05b}). These
models can easily be extended to incorporate indirect paths
of control vie the integrated model of  Sec. \ref{sec:nontop}.

Given any model for control, there is always a drawback in estimating
real-world control or power: shareholders do not only act as
individuals but can collaborate in shareholding coalitions and give
rise to so-called voting blocks. The theory of political voting games
in cooperative game theory has been applied to the problem of
shareholder voting in the form of so-called {\it power indices}
\cite{Leech02a}.  However, the employment of power indices for
measuring shareholder voting behavior has failed to find widespread
acceptance due to computational, inconsistency and conceptual issues
\cite{Leech02a}.

The so-called degree of control $\alpha$ was introduced in
\cite{CubbinLeech83} as a probabilistic voting model
measuring the degree of control of a block of large shareholdings as
the probability of it attracting majority support in a voting
game. Without going into details, the idea is as follows. Consider a
shareholder $i$ with ownership $W_{ij}$ in the stock $j$. Then the
control of $i$ depends not only on the value in absolute terms of
$W_{ij}$, but also on how dispersed the remaining shares are (measured
by the Herfindahl index). The more they tend to be dispersed, the
higher the value of $\alpha$. So even a shareholder with a small
$W_{ij}$ can obtain a high degree of control. The assumptions
underlying this probabilistic voting model correspond to those behind
the power indices. However, $\alpha$ suffers from drawbacks. It gives
a minimum cut-off value of $0.5$ (even for arbitrarily small
shareholdings) and hence Eq. (\ref{eq:sum_a}) is violated, meaning
that it cannot be utilized in an integrated model. The computation of
$\alpha$ can become intractable in situations with many shareholders.

To summarize, our measure of control extends existing integrated
models using fixed thresholds by incorporating insights from
probabilistic voting models (the analytical expressions of $H_{ij}$
and $\alpha$ share very similar behavior), and, furthermore, $\tilde
H_{ij}$ can be computed efficiently for large networks.

\end{appendix}

\bibliographystyle{plain}

\end{document}